\documentclass[natbib]{svjour3}
\usepackage{url}

\journalname{Space Science Reviews}

\ifx\pdftexversion\undefined
  \usepackage[dvips]{epsfig}
\else
  \usepackage[pdftex]{epsfig}
\fi

\graphicspath{{figs/},{./}}

\newcommand{\citen}[1]{\citeauthor{#1} \citeyear{#1}}

\newcommand{\apj}{{Astrophys.~J.}}

\newcommand{\solphys}{{Solar~Phys.}}

\newcommand{\grl}{{Geophys.~Res.~Lett.}}

\newcommand{\mnras}{{Monthly~Not.~Roy.~Astr.~Soc.}}

\def\ga{\mathrel{\mathchoice {\vcenter{\offinterlineskip\halign{\hfil
 $\displaystyle##$\hfil\cr>\cr\sim\cr}}}
 {\vcenter{\offinterlineskip\halign{\hfil$\textstyle##$\hfil\cr
 >\cr\sim\cr}}}
 {\vcenter{\offinterlineskip\halign{\hfil$\scriptstyle##$\hfil\cr
 >\cr\sim\cr}}}
 {\vcenter{\offinterlineskip\halign{\hfil$\scriptscriptstyle##$\hfil\cr
 >\cr\sim\cr}}}}}

\begin{document}



\title{The Magnetic Sun: Reversals and Long-Term Variations}

\author{K.~Petrovay
\and U. R. Christensen
}

\institute{
K. Petrovay
\at
E\"otv\"os University, Department of Astronomy, Budapest, Pf.~32,
H-1518 Hungary\\
           \and
U. R. Christensen
\at MPI f\"ur Sonnensystemforschung, D-37191 Katlenburg-Lindau, Germany
}

\date{Received: date / Accepted: date}

\authorrunning{Petrovay \& Christensen}
\titlerunning{The magnetic Sun}

\maketitle

\begin{abstract}
A didactic introduction to current thinking on some aspects of the solar dynamo
is given for geophysicists and planetary scientists.

\end{abstract}



\section{Introduction}

For a long time, solar dynamo theory was in an advantageous situation compared
to planetary dynamo theory. In the Sun, dynamo generated magnetic fields can be
directly measured on the boundary of the turbulent, conducting domain where
they are generated; and the timescales of their variations are short enough for
direct observational follow-up. Solar observations also put many constraints on
the motions in the convective zone, significantly restricting the otherwise
very wide range of admissible mean field dynamo models.

In the past 15 years, however, planetary dynamo theorists have turned the table.
Realistic numerical simulations of the complete geodynamo have been made
possible by the rapid increase in the available computing power. While the
parameter range for which such simulations are feasible is still far from
realistic, extrapolations based on the available results have allowed important
inferences on the behaviour of planetary dynamos
(\citen{Christensen:plan.dynamo.solar}, \citen{Petrovay:solar.planet.dynamos}).

In the light of these developments, learning from solar dynamo theory may have
lost much of its former appeal to geophysicists, especially as it has become
increasingly clear that the two classes of dynamos operate in fundamentally
different modes, under very different conditions. Yet, keeping track with
advance in the other field may not be without profit for either area. Even
though the overall mechanisms may be very different, there may be many elements
of each system where intriguing parallels exist.

A comprehensive review of solar dynamo theory is beyond the scope of the present
article; for this, we refer the reader to papers by \cite{Petrovay:SOLSPA},
\cite{Charbonneau:livingreview}, \cite{Solanki+:RPP} and
\cite{Jones+:dynamo.rev}.

Aside from reproducing the dipole dominance and other morphological traits
of the geomagnetic field,
two aspects of the geodynamo that are critical for judging the merits of its
models are field reversals and long-term variations: whether or not a model can
produce such effects in a way qualitatively, and if possible quantitatively
similar to the geological record, has become a testbed for geodynamo
simulations. It may thus be of special interest to review our current
understanding of the analogues of these phenomena in the Sun. This is the
purpose of the present paper.

In Section~2 we outline what solar observations suggest about the causes of
reversals, i.e.\ the Babcock--Leighton mechanism. Two questions naturally
arising from this discussion are given further attention in Sections 3 and 4.
Section~5 briefly discusses some issues related to long-term variations of solar
activity, while Section~6 concludes the paper, pointing out some interesting
parallel phenomena in solar and planetary dynamos.

\section{Reversals and global magnetism on the Sun}

Information about large scale flows in the solar convective envelope can place
important constraints on dynamo models. In the solar photosphere (the thin layer
where most of the visible radiation originates) these flows can be directly
detected by the tracking of individual features and by the Doppler effect. But
in recent decades {\it helioseismology} (a technique analoguous to terrestrial
seismology, see review by \citen{Howe:LRSP}) has also shed light on subsurface
flows. In particular, the internal {\it differential rotation} pattern is now
known in much of the solar interior. It is characterized by a marked latitudinal
differential rotation (faster equator, slower poles) throughout the convective
envelope, while the radiative interior rotates like a rigid body. A thin
transitional layer known as the {\it tachocline} separates the two regions. {\it
Meridional circulation,} on the other hand is currently only known in the
outermost part of the convective zone where it is directed from the equator
towards the poles. A return flow is obviously expected in deeper layers but this
has not yet been detected.

The invention of the magnetograph in 1959 marked a breakthrough in the study of
solar magnetism. The output of this instrument, the magnetogram, is basically an
intensity-coded map of the circular polarization over the solar disk. Circular
polarization in turn is due to the Zeeman effect and, in a rather wide range (up
to field strengths of $\sim 1\,$kG), it scales linearly with the line of sight
component of the magnetic field strength. Apart from the saturation at kilogauss
fields, then, a magnetogram is essentially a map of the line of sight magnetic
field strength over the solar disk in the photosphere. Conventionally, fields
with northern polarity are shown in white, while fields with southern polarity
are shown in black.

\begin{figure}
\begin{center}
\epsfig{figure=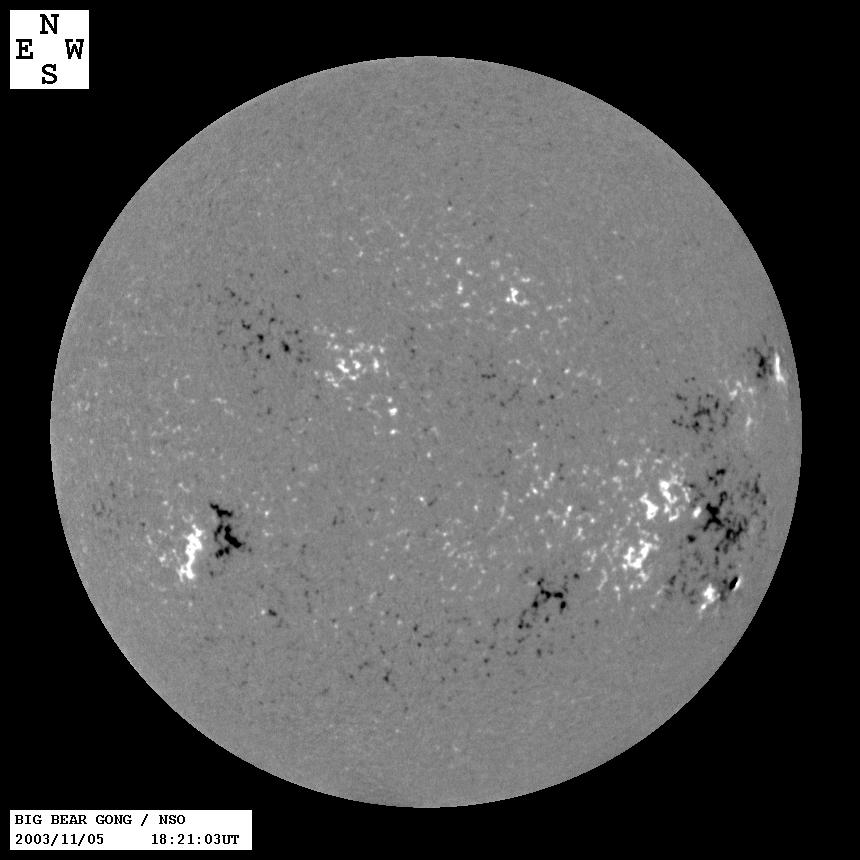, width=0.9\textwidth}
\end{center}
\caption{A solar magnetogram: a map of the line of sight magnetic field strength
over the solar disk. Fields with northern (southern) polarity are shown in white
(black).
\hfill\copyright Big Bear Observatory}
\end{figure}

Figure~1 shows an arbitrarily chosen magnetogram as an example. It is
immediately clear that the strongest magnetic concentrations, called {\it
active regions} (AR), occur in bipolar pairs. Filtergrams and non-optical
images showing the higher layers of the solar atmosphere confirm that these
pairs mark the footpoints of large magnetic loops protruding from the Sun's
interior into its atmosphere. In white light images these active regions appear
as dark {\it sunspot groups} and bright, filamentary {\it facular areas.} The
lifetime of spots, facul{\ae} and active regions is finite: turbulent motions in
the solar photosphere ultimately lead to their dispersal over a period not
exceeding a few weeks.

In addition to the strong active regions, Fig.~1 also displays some weaker and
more extended magnetic concentrations, also in bipolar pairs (e.g.\ a bit upper
left from the center). These features, only seen in magnetograms, are the
remains of decayed active regions, the bipolar pair of flux concentrations
being dispersed over an ever wider area of the solar surface. Ultimately, all
that is left is a pair of {\it unipolar areas} ---areas of quiet sun where
the ubiquitous small-scale background magnetic field is dominated by one
polarity or another.

\begin{figure}
\begin{center}
\epsfig{figure=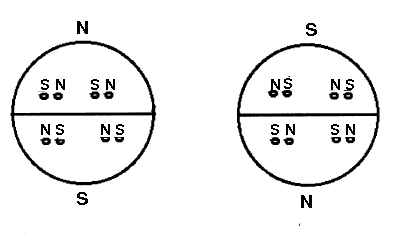, width=0.7\textwidth}
\end{center}
\caption{Sketch illustrating Hale's polarity rules in the growing phase of two
consecutive 11-year cycles (left and right, respectively)}
\end{figure}

We can see that the bipolar magnetic pairs are mostly oriented in the E-W
direction, (with a slight tilt, to be discussed below). In the northern
hemisphere, the N polarity patches lie to the west of their S polarity pairs,
while in the southern hemisphere the situation is opposite. In the image, the
direction of solar rotation is left to right and the rotational axis is
approximately vertical and in the plane of the sky. ``Western patches'' are
therefore referred to as the {\it preceding polarity} part of the active region,
while their eastern pairs are the {\it following polarity} part. The rule we
have just noticed then says that, at any given instant of time, preceding
polarities of solar active regions are uniform over one hemisphere and opposite
between the two hemispheres. Measurements performed during the course of
several solar cycles show that, in any given hemisphere, the preceding polarity
remains unchanged during the course of each 11-year solar cycle, while it
alternates between cycles. This regularity, known as {\it Hale's polarity
rules,} is schematically illustrated in Fig.~2. This implies that the
true period of solar activity is 22 years in the magnetic sense.

This sketch also shows one further regularity in the polarities. The weak
large-scale magnetic field is usually opposite near the two rotational poles,
and these polarities also alternate with a 22 year periodicity. However, the
phase of this 22 year cycle is offset by about $\pi/2$ from the active region
cycle, i.e. magnetic pole reversal does not occur in solar minimum. Instead,
reversals typically take place 1--2 years after solar maximum, right in the
middle of a cycle (as cycle profiles are asymmetric).

\begin{figure}
\epsfig{figure=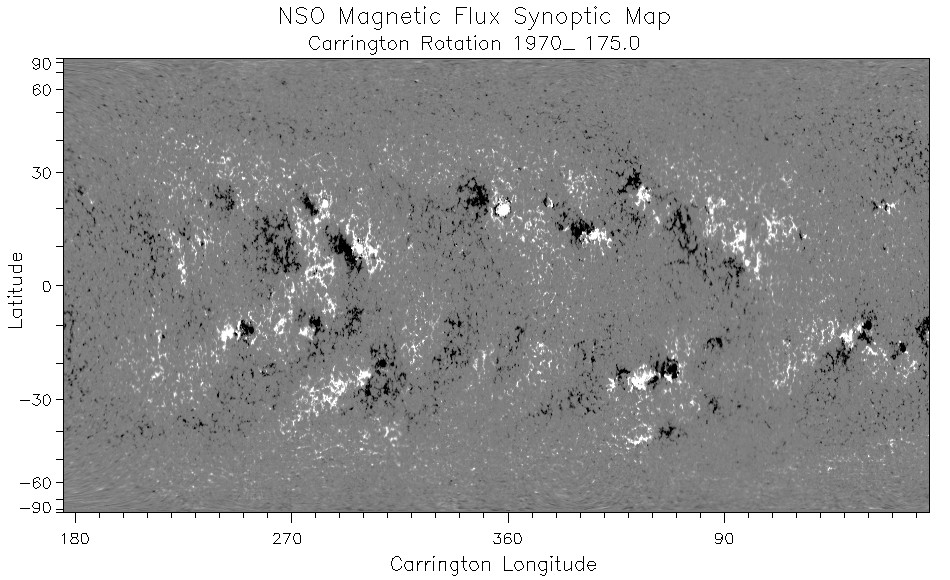, width=\textheight, angle=90}
\caption{A synoptic magnetic map of the Sun \hfill{\small\copyright NSO}}
\end{figure}

Systematic magnetograph studies coupled with pioneering work on magnetic flux
transport shed light on the apparent origin of the field reversal already in the
1960s (\citen{Babcock}, \citen{Leighton:diffusion}). For a better understanding
of how this so-called Babcock--Leighton mechanism works, consider a {\it
synoptic magnetic map} like the one shown in Fig.~3. Such synoptic maps are
essentially constructed by taking a narrow vertical strip from the center of
each daily magnetogram (i.e. along the central meridian) and sticking them
together from left to right, in a time sequence covering one synodic solar
rotation. In this way we arrive at a ``quasi-instantaneous'' (in as much as
$P_{\mathrm{rot}}\ll 11$ years) magnetic map of the full solar surface. In
Fig.~3 we see further ample evidence of Hale's polarity rules, but what is more
interesting is the systematic deviation of the orientation of bipolar pairs
from the E--W direction. It is clear that following polarity parts of active
regions are systematically closer to the poles than their preceding polarity
counterparts, and the tilt angle of the AR axis increases with heliographic
latitude ---a phenomenon known as {\it Joy's law.}

\begin{figure}
\epsfig{figure=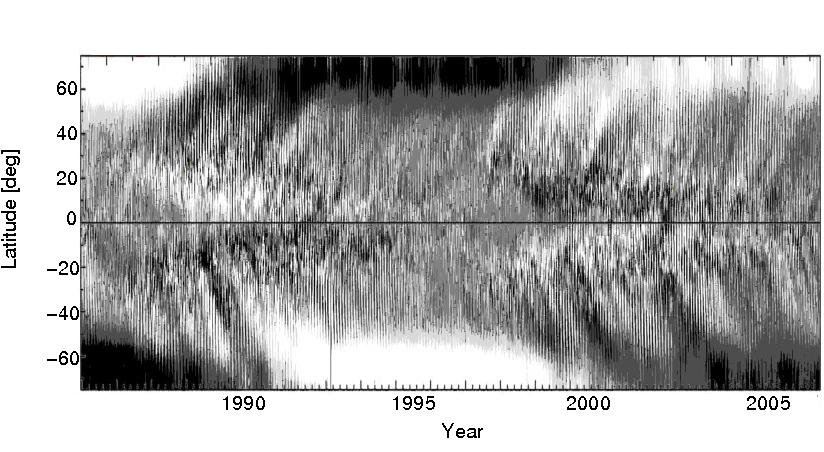, width=\textwidth}
\caption{The magnetic butterfly diagram: longitudinally averaged  
magnetic field strength as a function of latitude and time, based on Mt.~Wilson
Observatory data. The grayscale goes from -2 G (black) to +2 G (white).}
\end{figure}

As a consequence of Joy's law, upon the decay of active regions the resulting
following polarity unipolar areas will be predominantly found on the poleward
side of the active latitude belt. Now recall from Fig.~2 that in the early phase
of a solar cycle, the following polarity in a given hemisphere is opposite to
the polarity of the corresponding pole. As the decay of ever newer active
regions replenishes its magnetic flux, this following polarity belt, located
poleward of the active latitudes, will expand towards the pole, a process
greatly helped by turbulent magnetic diffusion and by the advection of the large
scale magnetic fields due to the Sun's large scale meridional circulation.
Ultimately, shortly after solar maximum, the preceding polarity patches around
the two poles shrink to oblivion, and following polarity takes over: a reversal
has taken place.

The process can be followed in the ``magnetic butterfly diagram'' shown in
Fig.~4. Each pixel in this plot represents the intensity-coded value of the
longitudinally averaged magnetic field strength at the given heliographic
latitude and time. The butterfly wing shaped areas of intense magnetic activity
at low latitudes represent the activity belts, migrating from medium latitudes
towards the equator during the course of each solar cycle. The poleward drift
apparent at high latitudes, in turn, corresponds to the poleward expansion of
following polarity areas discussed above. It is clear that the reversal of the
predominant magnetic polarity near the poles is due to this poleward drift,
confirming the Babcock--Leighton scenario.

However compellingly the observations argue for such a scenario, two open
questions remain.

1. What is the origin of the latitude-dependent tilt of active regions (Joy's
law)?\\
In the Babcock--Leighton scenario this tilt is clearly the ultimate cause of
flux reversals. $\alpha\omega$ dynamo models rather naturally lead to
predominantly toroidal, i.e.\ E--W oriented magnetic fields, so the tilt
compared to this prevalent direction is likely to develop during the process of
{\it magnetic flux emergence} from deeper layers into the atmosphere. This leads
us to discuss flux emergence models in Section~3.

2. Do magnetic flux redistribution processes seen at the surface indeed play an
important role in the dynamo, or are they just manifestations of similar,
more robust processes taking place deep within the Sun?

The nice consistent cause-and-effect chain involved in the Babcock--Leighton
scenario (AR tilt $+$ AR decay $\rightarrow$ $f$-polarity unipolar belt;
$f$-polarity belt $+$ flux transport $\rightarrow$ reversal) seems to argue
strongly for the first option. However, as we will see in Section~4 below,
keeping the surface physically decoupled from the deeper layers is not easy,
and such models invariably need to rely on some rather dubious physical
assumptions. Thus, while parametric models of the first type can be fine-tuned
to show impressive agreement with observations, their physical foundations
remain shaky. Conversely, models based on more sound and plausible physical
assumptions have had limited success in reproducing the details of
observations. This constitutes the main dichotomy in current (mainstream) solar
dynamo thinking, to be discussed in Section~4.

\begin{figure}
\epsfig{figure=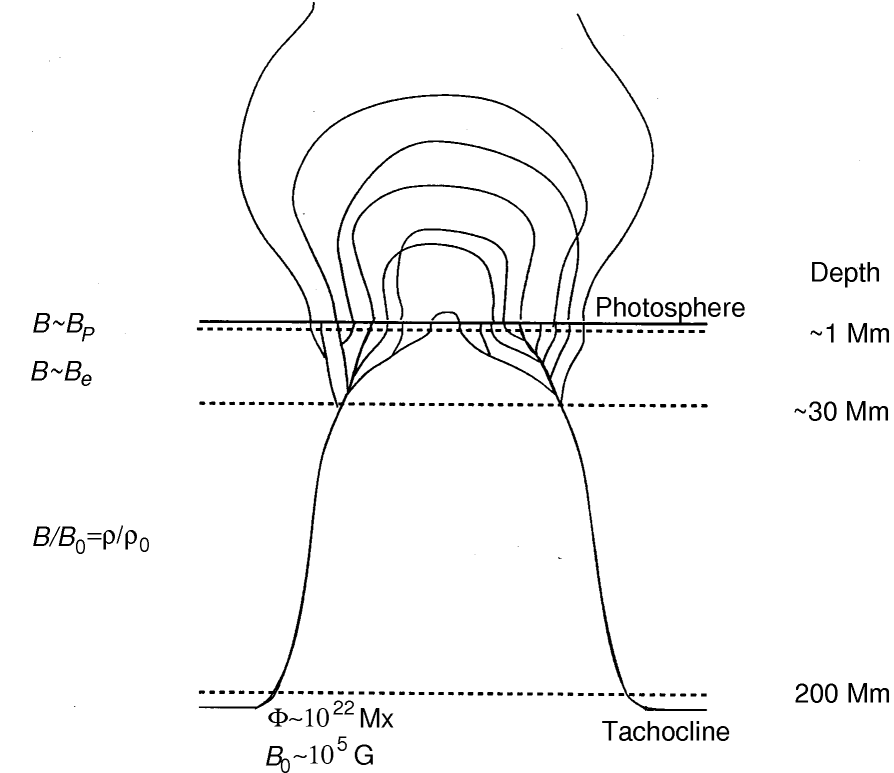, width=0.9\textwidth}
\caption{Sketch of the probable subsurface magnetic structure of an active
region. $B_P=(2\mu_0 P)^{1/2}$ is the pressure equipartition field strength, $P$
being the thermal gas pressure outside the flux tube. $B_e$
is the turbulent equipartition field strength as defined in the text.}
\end{figure}

\section {Flux emergence}

The current mainstream picture of the subsurface magnetic structure of solar
active regions is sketched in Fig.~5. The observed distribution and proper
motion patterns of sunspots and other magnetic elements in active regions are
strongly suggestive of the so-called {\it magnetic tree scenario:} the cluster
of smaller and larger magnetic flux tubes that manifest themselves as sunspots,
facular points etc. in the photospheric layers of an active region are all
connected in the deeper layers in a tree-like structure. The characteristic
sizes of unipolar patches in a typical bipolar active region suggest that the
trunk of this magnetic tree starts to fragment into branches at a relatively
shallow depth, on the order of $\sim 10\,$\% of the thickness of the convective
zone. During the emergence of the tree  structure, magnetic elements,
corresponding to mesh points of the tree branches with the surface, naturally
seem to converge into larger elements, ultimately forming large sunspots. This
is just how sunspots are observed to be formed.

The natural initial configuration for such a magnetic loop, fragmented in its
upper reaches, is a strong toroidal (i.e.\ horizontal and E--W oriented)
magnetic flux bundle lying below the surface.  The high field strength is a
requirement imposed by the strict adherence to Hale's polarity rules. Indeed,
there are almost no exceptions to these rules among large active regions, so
the drag force associated with the vigorous turbulent  convective flows in the
solar envelope must be small compared to other forces acting on the tube, in
particular the curvature force. As the total magnetic flux of the tube must be
similar to the flux of large active regions ($\sim 10^{22}\,$Mx), from this
condition it can be estimated that the field strength must significantly exceed
the equipartition field $B_e$  where the magnetic and turbulent pressures are
equal ($B_e^2/2\mu_o=\rho \langle v^2\rangle/2$). 
It follows that $B\ga 10^4\,$G ($\sim 1\,$T)
in the deep convective zone.

It is however well known (\citen{Parker:buoy.prob}) that such strong horizontal
magnetic flux tubes cannot be stably stored in a superadiabatic
or neutrally stable environment.
The magnetic pressure implies that the gas pressure inside the flux tube is
reduced in order to maintain pressure equilibrium with the environment. The
associated decrease in density of the compressible plasma makes the flux tube
magnetically buoyant. 
In order for such tubes to remain rooted in the solar interior while only a finite
section of them emerges in the form of a loop, the initial tube must reside
below the convective zone proper, in the subadiabatically stratified layer
below. This layer coincides with the region of strong rotational shear between
the rigidly rotating solar interior and the differentially rotating convective
envelope, the above mentioned tachocline. 


The emergence of flux loops formed on such strong toroidal flux bundles lying in
the tachocline is driven by an instability. Often referred to as the Parker
instability, this is a buoyancy driven instability of finite wavenumber
perturbations of the tube (while the $k=0$ mode remains stable, so the tube
remains rooted in the tachocline). The first models of this process employed the
thin flux tube approximation (\citen{Spruit:TFT}) where the tube is essentially
treated as a one-dimensional object. The emergence process was first followed
into the nonlinear regime, throughout the convective zone, by
\cite{FMI:classic}; the action of Coriolis force and spherical geometry were
incorporated in later models. Since the mid-1990s models have started to break
away from the thin flux tube approximation and by now full 3D simulations of the
emergence process have become the norm (see review by \citen{Fan:LRSP}). Indeed,
flux emergence models represent probably the most successful chapter in the
research of the origin of solar activity in the last few decades.

As the total magnetic flux of the toroidal flux tube is set equal to typical
active region fluxes, emergence models are left with only one free parameter:
the initial field strength $B_0$. (Twist is introduced as a further free
parameter in 3D models.) Comparing models with different values of $B_0$ to the
observed structure and dynamics of sunspot groups, it was recognized about two
decades ago (\citen{Petrovay:strongfield}) that the value of $B_0$ must be
surprisingly high, in the order of $10^5$ G, exceeding $B_e$ by an order of
magnitude. This came as a surprise, as flux expulsion in a turbulent medium was
known to concentrate diffuse magnetic fields in tubes and amplify them to $\sim
B_e$, but not to significantly higher values. Yet there are at least four
independent lines of argument to support this assertion.

\begin{figure}
\epsfig{figure=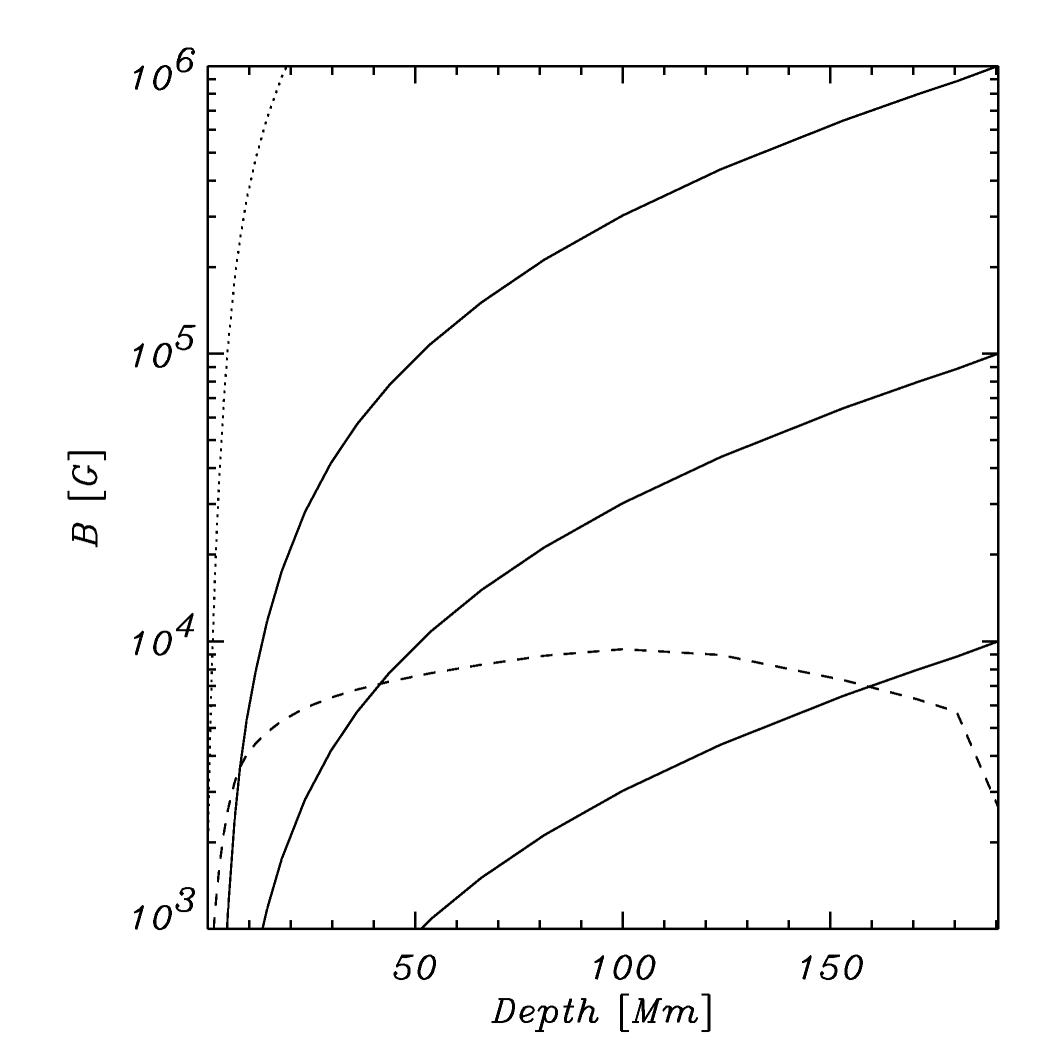, width=0.7\textwidth}
\caption{Magnetic field strength as a function of depth in the quasiadiabatic
convective zone in an emerging flux tube for initial field strenghts of 10, 100,
and 1000 kG (solid lines). Dashed line: turbulent equipartition field;
dotted line: thermal equipartition field}
\label{fig:fluxemerg}
\end{figure}

\begin{enumerate}

\item {\it Fragmentation depth.} As the rise through the convective zone takes
place on a relatively short timescale ($\sim 1\,$month), matter inside the flux
tube expands adiabatically. Most of the convective zone (except the uppermost
few hundred km) is also very nearly adiabatically stratified, and the magnetic
pressure is negligible compared to  the thermal pressure here, so the relative
density contrast between the inside of the tube and its surroundings will remain
relatively small. The field strength inside the loop is then given by
$B/B_0=\rho/\rho_0$, the '0' index referring to values at the bottom of the
convective zone. The resulting values of $B$ as a function of depth $z$ are
plotted in Fig.~\ref{fig:fluxemerg}. It is apparent that at a certain depth $B$
drops below the turbulent equipartition field strength $B_e$. Above this level
the magnetic field is unable to suppress turbulence and the external turbulent
motions will penetrate the tube. Flux expulsion processes taking place in
parallel with the further emergence of the loop are then expected to fragment
the top of the loop into a number of smaller flux tubes, resulting in the
magnetic tree structure. As we have seen, observations suggest that this
fragmentation occurs at a depth of a few tens of megameters:
Fig.~\ref{fig:fluxemerg} then clearly suggests $B_0\sim10^5\,$G.\\

\item {\it Emergence latitudes.}
For weaker tubes, the buoyancy and the curvature force are also weaker, so
the Coriolis force, independent of the field strength, will play a more dominant
role in the dynamics of the emerging loop. Tubes with $B_0\ll 10^5\,$G are so
strongly deflected by the Coriolis force that they will emerge approximately
parallel to the rotation axis. As the bottom of the convective zone lies at 0.7
photospheric radii, weak flux tubes emerging from here cannot reach the surface
at latitudes below 45 degrees, in contradiction to observations
(\citen{Choudhuri+Gilman}, \citen{Choudhuri:Coriolis}).\\

\item {\it Joy's law.} Tubes with higher values of $B_0$ will emerge
approximately radially, yet the effect of Coriolis force on them is not
negligible. The meridional component of the Coriolis force acting on the
downflows in the tilted legs of the emerging loop will twist the plane of the
loop out of the azimuthal plane, resulting in a tilt in the orientation of
active regions relative to the E--W direction. This tilt increases with
heliographic latitude, thus explaining the observed Joy's law. Quantitative
comparison between models and observations again shows that Joy's law is best
reproduced for $B_0\sim 10^5\,$G (\citen{D'Silva+Choudhuri}).\\

\begin{figure}
\epsfig{figure=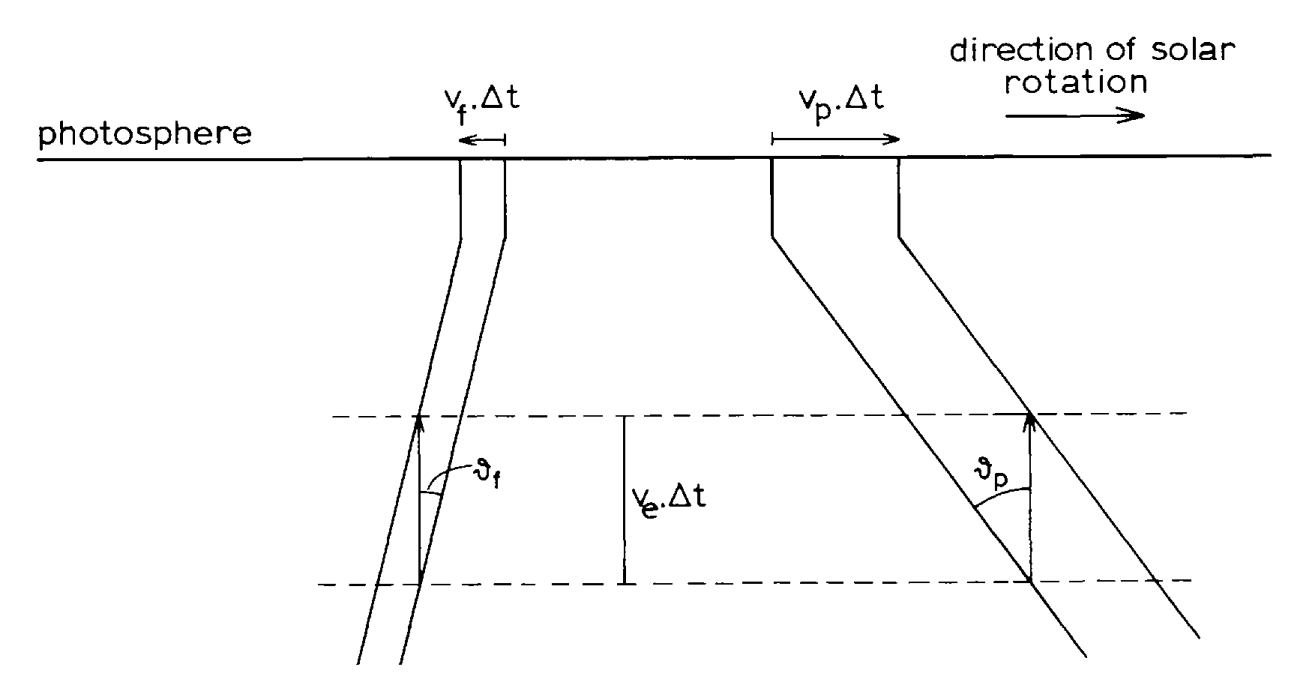, width=0.9\textwidth}
\caption{Interpretation of sunspot proper motions by the emergence of an
asymmetric flux loop (after \citen{vDG+Py:asym1})}
\label{fig:asymm}
\end{figure}

\item {\it Sunspot proper motions.} The azimuthal component of the Coriolis
force acting on the downflows in the tilted legs of the emerging loop has two
consequences. On the one hand, as this component is westward in both legs, it
will distort the shape of the emerging loop so that it will become
asymmetrical, the following leg being less inclined to the vertical than the
preceding leg. On the other hand, this westward force results in a wavelike
translational motion of the loop as a whole compared to the ambient medium: the
active region is then expected to rotate faster than quiet sun plasma. It is
indeed well known that sunspots and other magnetic tracers generally show some
superrotation compared to the Doppler rotation rate of the ambient plasma. In
the past, some confusion arose due to the fact that this superrotation appears
to be time dependent: newborn sunspot groups rotate fastest, their rotation
rate steadily declines during the growth phase of the group, until it becomes
stagnant at a rate only slightly above the plasma rotation rate from the time
when the group reaches its maximal development. This apparent change in the
rotation rate, however, was shown to be a purely geometrical projection effect
(\citen{vDG+Py:asym1}), as a consequence of the asymmetrical shape of the loop
(cf. Fig.~\ref{fig:asymm}). Again, detailed quantitative comparisons of sunspot
proper motion observations with the dynamics of emerging flux tube models
(\citen{FMI+:asym}, \citen{Caligari+:movie}) indicate optimal agreement for
$B_0\sim10^5\,$G.

\end{enumerate}

In summary: flux emergence models have led to the rather firm conclusion that
solar active regions are the product of the buoyancy driven rise of strong
magnetic flux loops through the convective zone. The loops arise from small
perturbations of strong toroidal flux bundles lying in the solar tachocline, at
the bottom of the convective zone. (These tubes may be preexistent, or
alternatively they may detach from a continuous flux distribution as a result
of the perturbation.) A number of independent arguments indicate that the
field strength in these toroidal tubes is on the order of $10^5\,$G. The
origin of the tilt in active regions orientations relative to the E--W
direction is clearly identified as the action of Coriolis force on the emerging
flux loops, twisting them out of the azimuthal plane.

\section{Alternative global scenarios for the solar dynamo}

As we have seen in Sections~2 and 3, the Babcock--Leighton mechanism offers a
very attractive explanation of the cyclic polar reversals and activity
variations observed on the Sun. During the flux emergence process, the Coriolis
force twists the plane of the flux loops out of the azimuthal plane, so they
acquire a poloidal magnetic field component. With the turbulent dispersal of the
active regions, this poloidal field component contributes to the large-scale
diffuse solar magnetic field. Advected towards the poles by meridional
circulation, it ultimately brings about the reversal of the global poloidal
magnetic field of the Sun. This reversed poloidal field is then advected down
into the tachocline by the meridional circulation near the poles. Continuity
requires that plasma advected to the poles near the surface by meridional
circulation must be returned to the equator at some depth; in the simplest case
of a one-cell circulation this will occur near the bottom of the convective
zone. This deep equatorward counterflow then advects the poloidal field towards
the equator at some slower speed, while differential rotation winds it up,
resulting in an ever stronger toroidal field component. By the time it reaches
latitudes below about $35^\circ$, this advected toroidal field reaches, at least
intermittently, the intensity of $\sim 10^5\,$G and starts to erupt in the form
of buoyancy driven loops, closing the cycle. The equatorward propagation of
active latitudes, manifest in the butterfly diagram, is thus solely due to the
advection of toroidal fields by the meridional circulation.

\begin{figure}
\begin{minipage}{0.33\textwidth}
{Flux transport dynamos:}\\

\epsfig{figure=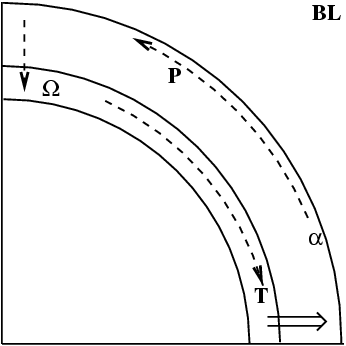,width=0.9\textwidth}

\end{minipage}
\begin{minipage}{0.3\textwidth}
\epsfig{figure=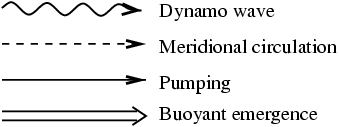,width=0.9\textwidth}
\end{minipage}
\begin{minipage}{0.33\textwidth}
{Interface dynamos:}\\

\epsfig{figure=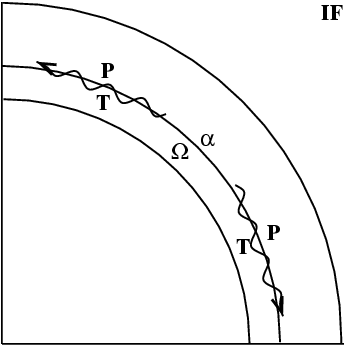,width=0.9\textwidth}

\end{minipage}
\caption{Sketches illustrating the basic concepts of the two main solar dynamo
scenarios}
\end{figure}

This scenario was already qualitatively outlined by \cite{Babcock} (with the
difference that he attributed the rise and tilt of flux loops to helical
convection). By now we have detailed quantitative models for the solar dynamo
along these lines (\citen{Dikpati+Charbonneau}, \citen{Nandy+Chodhuri:dynamo}).
As all migration phenomena in the butterfly diagram are explained by the
advection of magnetic flux in this picture, these models are known as {\it flux
transport dynamos.}

Flux transport models have become quite successful in reproducing many of the
observational details such as the shape of the butterfly diagram. This is to a
large extent due to their very good parametrizability. There are, however,
serious doubts concerning their physical consistency.

One issue concerns vertical flux transport. In order to have a poleward flux
transport near the surface and an equatorward transport near the bottom, the
surface and the tachocline must be kept {\it incomunicado} on a timescale
comparable to the solar cycle. However, simple mixing length estimates suggest
that the turbulent magnetic diffusivity in the convective zone is in the range
$10^2$--$10^3\,$km$^2$/s, so the timescale for the surface field to diffuse down
to the bottom across the the convective zone of depth $d=2\cdot 10^5\,$km is a
few years, certainly less than the solar cycle length. The above mentioned value
of the diffusivity is confirmed by calibrations based on surface flux
redistribution. So, in order to effectively decouple surface and bottom, flux
transport dynamos invariably need to rely to some ad hoc assumptions regarding
magnetic diffusivity: suppressing it in the bulk of the convective zone, making
it highly anisotropic etc.

A further difficulty is related to the flow pattern envisaged in these kinematic
models. The equatorward return flow of meridional circulation is assumed to
spatially overlap with the tachocline, i.e. the subadiabatic layer where the
toroidal field can be stably stored and where it is amplified by the strong
rotational shear. But this assumption is very dubious from a dynamical/thermal
point of view. To penetrate the subadiabatically stratified upper radiative
zone, the plasma partaking in the meridional flow would have to get rid of its
extra entropy so buoyancy will not inhibit its submergence. This, however, can
only occur on a slow, thermal timescale. Numerical estimates show that the
maximal circulation speed in the subadiabatic tachocline is a few cm/s, way too
slow for flux transport models to work.

An alternative  approach to the solar dynamo is, then, to try to construct
models from first principles, without introducing physically unsubstantiated
assumptions. Such a model cannot ignore the achievements of flux emergence
models and needs to be based on the assumption that the strong toroidal flux
tubes responsible for solar activity phenomena reside in the subadiabatic
tachocline, and then are presumably also generated there, given the strong
rotational shear. 

Significant work has been done on the role of a stably stratified convective
overshoot layer, coincident witha shear layer (tachocline) in large-scale
dynamos (cf.~\citen{Tobias:penet.rev} and \citen{Brandenburg:dynamo.rev}). The
slow meridional circulation allowed here cannot be responsible for the
latitudinal migration of the toroidal field as seen in the butterfly diagram.
Instead, this must be the manifestation of a classic dynamo wave, as already
suggested by \cite{Parker:cyclonic}. The tachocline is, however, probably not
turbulent enough to support a strong $\alpha$-effect, so the site of the
$\alpha$-effect must be in the convective zone, adjacent to the tachocline,
where convective downflows diverge, resulting in $\alpha<0$ in the northern
hemisphere as required for an equatorward propagating dynamo wave at the low
latitudes where $\partial\Omega/\partial r>0$.

In these  {\it interface dynamos}, then, the dynamo wave is excited as a surface
wave on the interface between the tachocline and the convective zone. Their
classic analytical prototype was constructed by \cite{Parker:interface}. At
higher latitudes, where $\partial\Omega/\partial r<0$ this model naturally
results in a poleward propagating dynamo wave, possibly explaining the poleward
migration of unipolar areas and other phenomena in this part of the Sun. This is
an attractive feature of these models, but the question naturally arises,
whether the equatorward return flow that must be present in the lower convective
zone if not in the tachocline, i.e.\ on one side of the interface will affect
these results. This was examined by \cite{Petrovay+Kerekes:intflow} in an
extension of Parker's analytical work to the case of a meridional flow. It was
found that for parameters relevant to the solar case the meridional flow is
unable to overturn the direction of propagation of dynamo waves, nor will it
significantly affect their growth rates. For a related study of the effect of
other flux transport mechanisms on the interface dynamos see 
\cite{Mason+:interface.fluxtransp}.

A number of detailed numerical interface dynamos models have been constructed
for the Sun (\citen{Charbonneau+McGregor:IFdynamo}, \citen{Markiel+Thomas},
\citen{Dikpati+:interface}). It must be admitted that at present they are unable
to reproduce the observed features of the solar activity cycle as satisfactorily
as some flux transport models can. However, this is clearly a consequence of the
fact that these models lack the kind of physical arbitrariness that
characterizes some flux kinematical transport models, where the arbitrary
prescription of flow geometries and amplitudes (esp.\ for the meridional flow)
leaves much more space to play around with parameters until an acceptable fit to
observations results.

\section{Long term variations}

In the geodynamo, field reversals and long-term variations are closely related.
Indeed, reversals are the most important kind of long term variation. In the
Sun, however, the field regularly reverses in every 11-year cycle: in fact,
reversals are the essence of the 11/22-year cycle.  Long term variations, in
turn, are a completely independent topic. It is impossible to give a
comprehensive review of this area in the present introductory paper  (see
\citen{Usoskin+Mursula:review}, \citen{Usoskin:LRSP} for much more exhaustive
overviews); instead, we limit ourselves to mentioning some salient points and
recent results.

\begin{figure}
\epsfig{figure=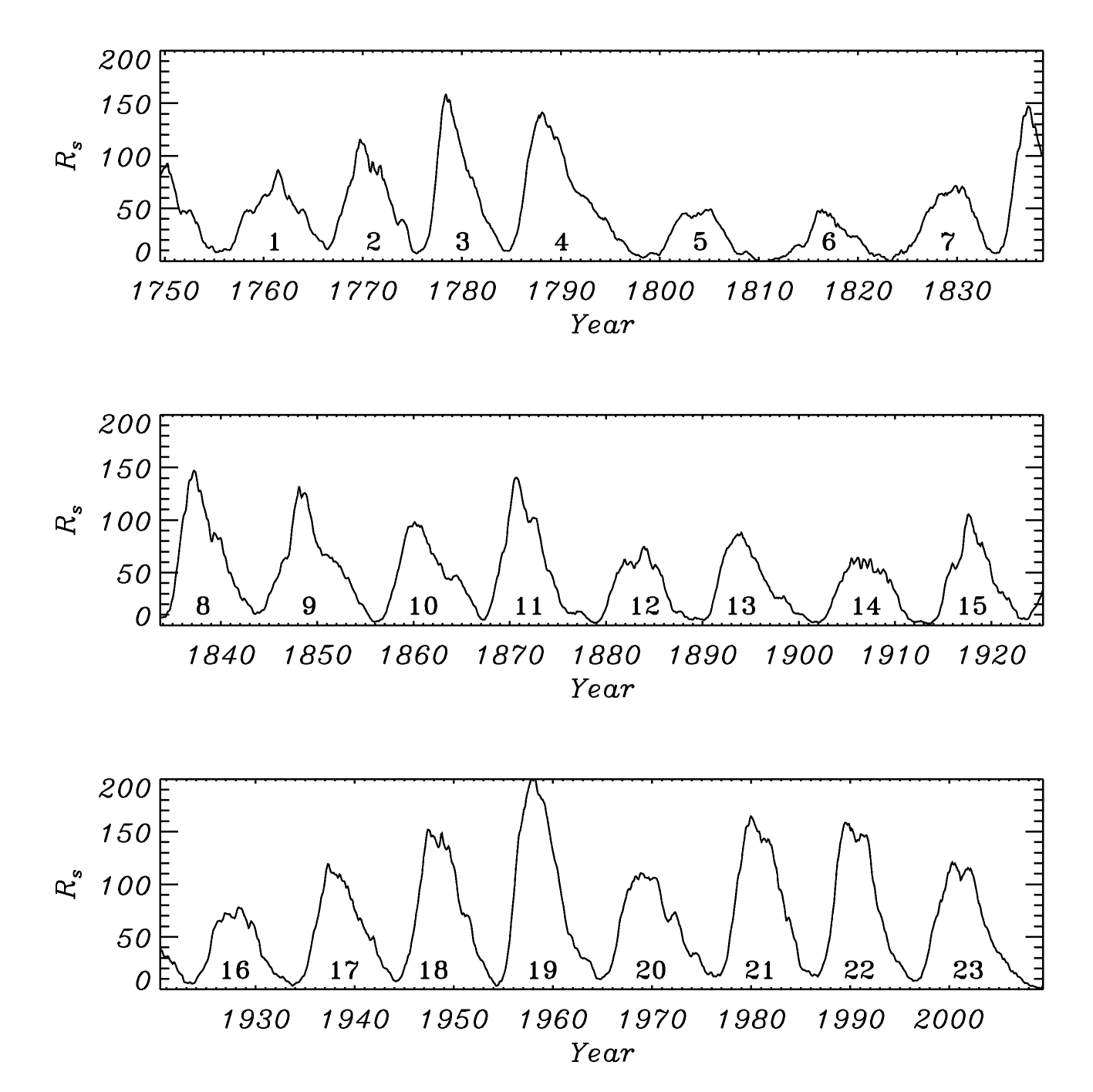, width=0.9\textwidth}
\caption{Variation of the smoothed (13-month boxcar average) relative sunspot
number with time. Solar cycles are numbered.}
\label{fig:spotcycle}
\end{figure}

Figure~\ref{fig:spotcycle} presents the variation of the relative sunspot
number in the last three centuries, for which the most reliable direct data
exist. From historical solar observations going back another century we know
that immediately before the period covered by Fig.~\ref{fig:spotcycle}, the Sun
underwent an unusally quiet period lasting nearly 70 years: in this so-called
Maunder minimum there were hardly any spots seen on the Sun at all. At the
other extreme lies the Modern Maximum: the series of unusually strong cycles
that started in the mid-20th century and seems to be just about to come to an
end now.

Various terrestrial proxies of solar activity have made it possible to
reconstruct long term activity variations (albeit not necessarily individual
cycles) over a substantially longer time span. The best results were found from
studies of the abundance of the cosmogenic $^{10}$Be isotope in Greenland ice
cores. These studies led to a reconstruction of the history of solar activity
in the last 9000 years with a time resolution of about 10 years
(\citen{Beer:SSRreview}). Recently, data with even better, annual resolution
have become available for the last six centuries
(\citen{Berggren:600yr.annual}).

One interesting result from these studies concerns the histogram of decadal
activity levels (\citen{Usoskin+:longterm.histogram}). It was found that the
overall shape of this histogram is compatible with a normal distribution;
however, significant excesses or ``shoulders'' appear at both extremes. In
other words, times of very low and very high solar activity are overabundant
relative to a Gaussian statistics. These results suggest that grand minima and
grand maxima are more than random fluctuations: they are indeed  physically
distinct states of the dynamo. Long term variations in the intermediate states,
in turn, seem to be driven by stochastic effects, resulting in nearly Gaussian
statistics.

Attempts to explain the long term variation in solar activity, as well as grand
minima and maxima, include stochastic fluctuations in dynamo parameters
(\citen{Ossendrijver2000}, \citen{Moss+:stochastic.dynamo}); nonlinear dynamos
with chaotic behaviour (\citen{Beer+:grand.minima},
\citen{Brooke+:grand.minima}) or with two alternative stationary dynamo
solutions (\citen{Petrovay:bimodal}). The shape of the histogram discussed
above suggests a bimodal solution with strong stochastic forcing resulting in
an actual solution that random walks in between the attractors for most of the
time.

It should be noted that radionuclide based reconstructions of solar activity
involve several uncertainties. On the long timescales we are concerned with,
the most important such effect is the parallel long term variation in the
geomagnetic field. The atmospheric cosmic ray flux, and thereby the
radionuclide production rate, is the net result of the shading effects of both
the interplanetary and the terrestrial magnetic fields, so long-term
normalization of the solar modulation crucially depends on the correct
subtraction of the geomagnetic contribution. Issues such as just how
exceptionally strong the most recent grand maximum (the Modern Maximum) was,
are strongly influenced by these effects.

\section{Conclusion: analogies and differences with respect to the geodynamo}

At the most basic level, the same processes generate magnetic field in
the Earth's core and inside the Sun: the shearing of magnetic field lines
by differential rotation ($\Omega$-effect) and their twisting by helical
motions ($\alpha$-effect) that obtain a preferred handedness from the action
of Coriolis forces. Also, some magnetic phenomena may have similar causes.
For example, paired flux spots at low latitudes in the geomagnetic field at
the core-mantle boundary have been tentatively explained by the expulsion of toroidal
flux tubes (\citen{Bloxham86}), in analogy to the generation of sunspots.
But although this interpretation is supported by some geodynamo models
(\citen{ChristensenOlson03}), other explanations for low-latitude magnetic
structures have been put forward (\citen{FinlayJackson03}). 
Meridional flow is thought to play the essential role for reversals of
the solar magnetic field. This has also been demonstrated in a simple 
reversing geodynamo model (\citen{WichtOlson04}). However, the reversal behaviour 
in this model is nearly cyclic as in case of the Sun, in contrast to the 
stochastic reversals of the geomagnetic field. In a less idealized geodynamo model with
random reversals, impulsive upwellings have been identified as the cause
for polarity changes (\citen{Aubert_etal08}). These upwellings transport and 
amplify a multipolar magnetic field from depth to the outer boundary. 
While possible analogies 
between the solar dynamo and the geodynamo can stimulate our thinking, we
must keep their limitations in mind.

Which differences in physical conditions lead to the rather distinct 
behaviour of the solar dynamo and the geodynamo? 
One important difference is that the plasma in the
solar convection zone is sufficiently compressible and that the 
field strength is high enough so that magnetic pressure and magnetic buoyancy
play an essential role for the dynamics of flux tubes. These effects are 
probably unimportant in Earth's core. Another difference is that the Coriolis
force has a stronger influence in the geodynamo than it has in the slowly
rotating Sun. This notion is supported by the observation that much more
rapidly rotating stars of low mass seem to have strong large-scale 
magnetic fields that are frequently dominated by the axial dipole component
(\citen{Donati_etal08}). Their observed field strengths 
follow the same scaling law as the observed fields of Earth and Jupiter and
the field intensity found in geodynamo models at sufficiently rapid rotation
(\citen{Christensen_etal09}).
A third difference arises from the much slower motion and therefore lower magnetic
Reynolds number in the geodynamo.

The moderate value of the magnetic Reynolds number makes the magnetic induction
process in the Earth's core amenable to direct numerical simulations without
the need to take recourse to turbulent magnetic diffusivities or parameterized
turbulent $\alpha$-effects. This is perhaps the most important reason for the
success of geodynamo models in reproducing many observed properties of the
geomagnetic field without need for ad-hoc assumptions. However,
our more limited knowledge of the geomagnetic field
at the top of the core, in comparison to that of the field in the solar
photosphere, makes the task simpler for a geodynamo modeller. Also,
helioseismology has revealed the distribution of zonal flow
in the solar convection zone and a fully consistent solar dynamo model must
reproduce this flow pattern as well as the magnetic field properties.
Comparable information is lacking for the Earth and a geodynamo model can be 
declared successful once it captures the general properties of the large-scale
geomagnetic field.

\begin{acknowledgements}
K. Petrovay's work on this review was supported by the Hungarian
Science Research Fund (OTKA) under grant no.\ K67746 and
by  the European Commission through the RTN programme SOLAIRE (contract
MRTN-CT-2006-035484).\\

\end{acknowledgements}


\begin{thebibliography}{}

\bibitem[\protect\citeauthoryear{Aubert, Aurnou and
  Wicht}{2008}]{Aubert_etal08}
J. Aubert, J. Aurnou, J. Wicht,
\newblock The magnetic structure of convection-driven numerical dynamos.
\newblock {\rm Geophys. J. Int.\/} {\bf 172}, 945 (2008)

\bibitem[\protect\citeauthoryear{{Babcock}}{1961}]{Babcock}
H.~W. {Babcock},
\newblock {The topology of the Sun's magnetic field and the 22-year cycle}.
\newblock {\rm \apj\/} {\bf 133}, 572 (1961)

\bibitem[\protect\citeauthoryear{{Beer}, {Tobias} and
  {Weiss}}{1998}]{Beer+:grand.minima}
J. {Beer}, S. {Tobias}, N. {Weiss},
\newblock {An active Sun throughout the Maunder minimum}.
\newblock {\rm \solphys\/} {\bf 181}, 237 (1998)

\bibitem[\protect\citeauthoryear{{Beer}, {Vonmoos} and
  {Muscheler}}{2006}]{Beer:SSRreview}
J. {Beer}, M. {Vonmoos}, R. {Muscheler},
\newblock {Solar variability over the past several millennia}.
\newblock {\rm Space Science Reviews\/} {\bf 125}, 67 (2006)

\bibitem[\protect\citeauthoryear{{Berggren}
  {et~al.}}{2009}]{Berggren:600yr.annual}
A.-M. {Berggren}, J. {Beer}, G. {Possnert}, A. {Aldahan}, P. {Kubik}, M.
  {Christl}, S.~J. {Johnsen}, J. {Abreu}, B.~M. {Vinther},
\newblock {A 600-year annual $^{10}$Be record from the NGRIP ice core,
  Greenland}.
\newblock {\rm \grl\/} {\bf 36}, 11801 (2009)

\bibitem[\protect\citeauthoryear{Bloxham}{1986}]{Bloxham86}
J. Bloxham,
\newblock The expulsion of magnetic flux from the {Earth's} core.
\newblock {\rm Proc. R. Soc. Lond.\/} {\bf A 87}, 669 (1986)

\bibitem[\protect\citeauthoryear{{Brandenburg}}{2009}]{Brandenburg:dynamo.rev}
A. {Brandenburg},
\newblock {Advances in Theory and Simulations of Large-Scale Dynamos}.
\newblock {\rm Space Sci.~Rev.\/} {\bf 144}, 87 (2009)

\bibitem[\protect\citeauthoryear{{Brooke}
  {et~al.}}{1998}]{Brooke+:grand.minima}
J.~M. {Brooke}, J. {Pelt}, R. {Tavakol}, A. {Tworkowski},
\newblock {Grand minima and equatorial symmetry breaking in axisymmetric dynamo
  models}.
\newblock {\rm {\aa}p\/} {\bf 332}, 339 (1998)

\bibitem[\protect\citeauthoryear{Caligari, {Moreno-Insertis} and
  Sch{\"u}ssler}{1995}]{Caligari+:movie}
P. Caligari, F. {Moreno-Insertis}, M. Sch{\"u}ssler,
\newblock {Emerging flux tubes in the solar convective zone I. Asymmetry, Tilt,
  and Emergence Latitude}.
\newblock {\rm \apj\/} {\bf 441}, 886 (1995)

\bibitem[\protect\citeauthoryear{Charbonneau}{2005}]{Charbonneau:livingreview}
P. Charbonneau,
\newblock {Dynamo models of the solar cycle}.
\newblock {\rm Living~Rev.~Sol.~Phys.\/} {\bf 2}, 2 (2005)

\bibitem[\protect\citeauthoryear{Charbonneau and
  MacGregor}{1997}]{Charbonneau+McGregor:IFdynamo}
P. Charbonneau, K.~B. MacGregor,
\newblock {Solar interface dynamos II. Linear, kinematic models in spherical
  geometry}.
\newblock {\rm \apj\/} {\bf 486}, 502 (1997)

\bibitem[\protect\citeauthoryear{{Choudhuri}}{1989}]{Choudhuri:Coriolis}
A.~R. {Choudhuri},
\newblock {The evolution of loop structures in flux rings within the solar
  convection zone}.
\newblock {\rm \solphys\/} {\bf 123}, 217 (1989)

\bibitem[\protect\citeauthoryear{{Choudhuri} and
  {Gilman}}{1987}]{Choudhuri+Gilman}
A.~R. {Choudhuri}, P.~A. {Gilman},
\newblock {The influence of the Coriolis force on flux tubes rising through the
  solar convection zone}.
\newblock {\rm \apj\/} {\bf 316}, 788 (1987)

\bibitem[\protect\citeauthoryear{Christensen, Holzwarth and
  Reiners}{2009}]{Christensen_etal09}
U.~R. Christensen, V. Holzwarth, A. Reiners,
\newblock Energy flux determines magnetic field strength of planets and stars.
\newblock {\rm Nature\/} {\bf 457}, 167 (2009)

\bibitem[\protect\citeauthoryear{Christensen and
  Olson}{2003}]{ChristensenOlson03}
U.~R. Christensen, P. Olson,
\newblock Secular variation in numerical geodynamo models with lateral
  variations of boundary heat flow.
\newblock {\rm Phys. Earth Planet. Inter.\/} {\bf 138}, 39 (2003)

\bibitem[\protect\citeauthoryear{{Christensen}, {Schmitt} and
  {Rempel}}{2009}]{Christensen:plan.dynamo.solar}
U.~R. {Christensen}, D. {Schmitt}, M. {Rempel},
\newblock {Planetary dynamos from a solar perspective}.
\newblock {\rm Space Sci.~Rev.\/} {\bf 144}, 105 (2009)

\bibitem[\protect\citeauthoryear{Dikpati and
  Charbonneau}{1999}]{Dikpati+Charbonneau}
M. Dikpati, P. Charbonneau,
\newblock {A Babcock--Leighton flux transport dynamo with solar-like
  differential rotation}.
\newblock {\rm \apj\/} {\bf 518}, 508 (1999)

\bibitem[\protect\citeauthoryear{{Dikpati}, {Gilman} and
  {MacGregor}}{2005}]{Dikpati+:interface}
M. {Dikpati}, P.~A. {Gilman}, K.~B. {MacGregor},
\newblock {Constraints on the applicability of an interface dynamo to the Sun}.
\newblock {\rm \apj\/} {\bf 631}, 647 (2005)

\bibitem[\protect\citeauthoryear{Donati {et~al.}}{2008}]{Donati_etal08}
J.-F. Donati, J. Morin, P. Petit, X. Delfosse, T. Forveille, M. Auri\`ere, R.
  Cabanac, B. Dintrans, R. Fares, T. Gastine, M.~M. Jardine, F. Ligni\`eres, F.
  Paletou, J.~C. Ramirez~Velez, S. Th\'eado,
\newblock Large-scale magnetic topologies of early m-dwarfs.
\newblock {\rm Mon. Not. R. Astron. Soc.\/} {\bf 390}, 545 (2008)

\bibitem[\protect\citeauthoryear{D'Silva and
  Choudhuri}{1993}]{D'Silva+Choudhuri}
S. D'Silva, A.~R. Choudhuri,
\newblock A theoretical model for tilts of bipolar magnetic regions.
\newblock {\rm {\aa}p\/} {\bf 272}, 621 (1993)

\bibitem[\protect\citeauthoryear{{Fan}}{2004}]{Fan:LRSP}
Y. {Fan},
\newblock {Magnetic fields in the solar convection zone}.
\newblock {\rm Living Rev. Sol. Phys.\/} {\bf 1}, 1 (2004)

\bibitem[\protect\citeauthoryear{Finlay and Jackson}{2003}]{FinlayJackson03}
C.~C. Finlay, A. Jackson,
\newblock Equatorially dominated magnetic field change at the surface of
  {Earth's} core.
\newblock {\rm Science\/} {\bf 300}, 2084 (2003)

\bibitem[\protect\citeauthoryear{{Howe}}{2009}]{Howe:LRSP}
R. {Howe},
\newblock {Solar Interior Rotation and its Variation}.
\newblock {\rm Living Rev. Sol. Phys.\/} {\bf 6}, 1 (2009)

\bibitem[\protect\citeauthoryear{{Jones}, {Thompson} and
  {Tobias}}{2009}]{Jones+:dynamo.rev}
C.~A. {Jones}, M.~J. {Thompson}, S.~M. {Tobias},
\newblock {The Solar Dynamo},
\newblock  {\rm Space Sci.~Rev.,}   in press (2009)

\bibitem[\protect\citeauthoryear{{Leighton}}{1964}]{Leighton:diffusion}
R.~B. {Leighton},
\newblock {Transport of magnetic fields on the Sun}.
\newblock {\rm \apj\/} {\bf 140}, 1547 (1964)

\bibitem[\protect\citeauthoryear{Markiel and Thomas}{1999}]{Markiel+Thomas}
J.~A. Markiel, J.~H. Thomas,
\newblock Solar interface dynamo models with a realistic rotation profile.
\newblock {\rm \apj\/} {\bf 523}, 827 (1999)

\bibitem[\protect\citeauthoryear{{Mason}, {Hughes} and
  {Tobias}}{2008}]{Mason+:interface.fluxtransp}
J. {Mason}, D.~W. {Hughes}, S.~M. {Tobias},
\newblock {The effects of flux transport on interface dynamos}.
\newblock {\rm \mnras\/} {\bf 391}, 467 (2008)

\bibitem[\protect\citeauthoryear{{Moreno-Insertis}}{1986}]{FMI:classic}
F. {Moreno-Insertis},
\newblock Nonlinear time-evolution of kink-unstable magnetic flux tubes in the
  convective zone of the sun.
\newblock {\rm {\aa}p\/} {\bf 166}, 291 (1986)

\bibitem[\protect\citeauthoryear{{Moreno-Insertis}, {Caligari} and
  {Schuessler}}{1994}]{FMI+:asym}
F. {Moreno-Insertis}, P. {Caligari}, M. {Schuessler},
\newblock {Active region asymmetry as a result of the rise of magnetic flux
  tubes}.
\newblock {\rm \solphys\/} {\bf 153}, 449 (1994)

\bibitem[\protect\citeauthoryear{{Moss}
  {et~al.}}{2008}]{Moss+:stochastic.dynamo}
D. {Moss}, D. {Sokoloff}, I. {Usoskin}, V. {Tutubalin},
\newblock {Solar grand minima and random fluctuations in dynamo parameters}.
\newblock {\rm \solphys\/} {\bf 250}, 221 (2008)

\bibitem[\protect\citeauthoryear{{Nandy} and
  {Choudhuri}}{2001}]{Nandy+Chodhuri:dynamo}
D. {Nandy}, A.~R. {Choudhuri},
\newblock {Toward a mean field formulation of the Babcock--Leighton type solar
  dynamo. I. {$\alpha$}-Coefficient versus Durney's Double-Ring Approach}.
\newblock {\rm \apj\/} {\bf 551}, 576 (2001)

\bibitem[\protect\citeauthoryear{{Ossendrijver}}{2000}]{Ossendrijver2000}
M.~A.~J.~H. {Ossendrijver},
\newblock {Grand minima in a buoyancy-driven solar dynamo}.
\newblock {\rm {\aa}p\/} {\bf 359}, 364 (2000)

\bibitem[\protect\citeauthoryear{{Parker}}{1955}]{Parker:cyclonic}
E.~N. {Parker},
\newblock {Hydromagnetic dynamo models}.
\newblock {\rm \apj\/} {\bf 122}, 293 (1955)

\bibitem[\protect\citeauthoryear{Parker}{1975}]{Parker:buoy.prob}
E.~N. Parker,
\newblock {The generation of magnetic fields in astrophysical bodies X:
  Magnetic buoyancy and the solar dynamo}.
\newblock {\rm \apj\/} {\bf 198}, 205 (1975)

\bibitem[\protect\citeauthoryear{{Parker}}{1993}]{Parker:interface}
E.~N. {Parker},
\newblock {A solar dynamo surface wave at the interface between convection and
  nonuniform rotation}.
\newblock {\rm \apj\/} {\bf 408}, 707 (1993)

\bibitem[\protect\citeauthoryear{{Petrovay}}{1991}]{Petrovay:strongfield}
K. {Petrovay},
\newblock {On the properties of toroidal flux tubes in the solar dynamo}.
\newblock {\rm \solphys\/} {\bf 134}, 407 (1991)

\bibitem[\protect\citeauthoryear{Petrovay}{2000}]{Petrovay:SOLSPA}
K. Petrovay,
\newblock {What makes the Sun tick?},
\newblock in {\em The solar cycle and terrestrial climate\/}, ESA Publ.~{\bf
  {SP-463}}, ~3 (2000)

\bibitem[\protect\citeauthoryear{{Petrovay}}{2007}]{Petrovay:bimodal}
K. {Petrovay},
\newblock {On the possibility of a bimodal solar dynamo}.
\newblock {\rm Astron.~Nachr.\/} {\bf 328}, 777 (2007)

\bibitem[\protect\citeauthoryear{{Petrovay}}{2009}]{Petrovay:solar.planet.dyna%
mos}
K. {Petrovay},
\newblock {Solar and planetary dynamos: comparison and recent developments},
\newblock in {\em Universal heliophysical processes\/}, ed.~by N. {Gopalswamy},
  D.~F. {Webb}, IAU Symp.~{\bf 257}, ~71 (2009)

\bibitem[\protect\citeauthoryear{{Petrovay} and
  {Kerekes}}{2004}]{Petrovay+Kerekes:intflow}
K. {Petrovay}, A. {Kerekes},
\newblock {The effect of a meridional flow on Parker's interface dynamo}.
\newblock {\rm \mnras\/} {\bf 351}, L59 (2004)

\bibitem[\protect\citeauthoryear{Solanki, Inhester and
  Sc{h{\"u}s}sler}{2006}]{Solanki+:RPP}
S. Solanki, B. Inhester, M. Sc{h{\"u}s}sler,
\newblock The solar magnetic field.
\newblock {\rm Rep.~Prog.~Phys.\/} {\bf 69}, 563 (2006)

\bibitem[\protect\citeauthoryear{{Spruit}}{1981}]{Spruit:TFT}
H.~C. {Spruit},
\newblock {Motion of magnetic flux tubes in the solar convection zone and
  chromosphere}.
\newblock {\rm {\aa}p\/} {\bf 98}, 155 (1981)

\bibitem[\protect\citeauthoryear{{Tobias}}{2009}]{Tobias:penet.rev}
S.~M. {Tobias},
\newblock {The Solar Dynamo: The Role of Penetration, Rotation and Shear on
  Convective Dynamos}.
\newblock {\rm Space Sci.~Rev.\/} {\bf 144}, 77 (2009)

\bibitem[\protect\citeauthoryear{{Usoskin}}{2008}]{Usoskin:LRSP}
I.~G. {Usoskin},
\newblock {A history of solar activity over millennia}.
\newblock {\rm Living Rev. Sol. Phys.\/} {\bf 5}, 3 (2008)

\bibitem[\protect\citeauthoryear{{Usoskin} and
  {Mursula}}{2003}]{Usoskin+Mursula:review}
I.~G. {Usoskin}, K. {Mursula},
\newblock {Long-term solar cycle evolution: review of recent developments}.
\newblock {\rm \solphys\/} {\bf 218}, 319 (2003)

\bibitem[\protect\citeauthoryear{{Usoskin}, {Solanki} and
  {Kovaltsov}}{2007}]{Usoskin+:longterm.histogram}
I.~G. {Usoskin}, S.~K. {Solanki}, G.~A. {Kovaltsov},
\newblock {Grand minima and maxima of solar activity: new observational
  constraints}.
\newblock {\rm {\aa}p\/} {\bf 471}, 301 (2007)

\bibitem[\protect\citeauthoryear{{van Driel-Gesztelyi} and
  {Petrovay}}{1990}]{vDG+Py:asym1}
L. {van Driel-Gesztelyi}, K. {Petrovay},
\newblock {Asymmetric flux loops in active regions, I}.
\newblock {\rm \solphys\/} {\bf 126}, 285 (1990)

\bibitem[\protect\citeauthoryear{Wicht and Olson}{2004}]{WichtOlson04}
J. Wicht, P. Olson,
\newblock A detailed study of the polarity reversal mechanism in a numerical
  dynamo model.
\newblock {\rm Geochem., Geophys,. Geosyst.\/} {\bf 5},
  doi:10.1029/2003GC000602 (2004)

\end{thebibliography}


\end{document}